\documentclass[aps,prl,twocolumn,groupedaddress,showpacs]{revtex4}
\usepackage{epsfig}
\usepackage{graphicx}% Include figure files
\newcommand{\la}{\langle}
\newcommand{\ra}{\rangle}
\newcommand{\up}{\uparrow}
\newcommand{\dn}{\downarrow}

\begin{document}

\title{Role of electron correlations in transport through domain walls in magnetic nanowires}
\author{M. A. N. Ara\'ujo$^{a,b}$\footnote{On leave from Departamento de F\'{\i}sica,
Universidade de \'Evora, P-7000-671, \'Evora, Portugal},
V. K. Dugaev$^{b,c,d}$, V. R. Vieira$^{b}$, J. Berakdar$^{d}$, and J. Barna\'s$^{e,f}$}
\affiliation{$^{a}$Department of Physics, Massachusetts Institute of Technology,
Cambridge MA 02139, U.S.A.}
\affiliation{$^{b}$CFIF and Departamento de F\'isica, Instituto Superior
T\'ecnico, Av. Rovisco Pais, 1049-001 Lisbon, Portugal}
\affiliation{$^c$Frantsevich Institute for Problems of Materials Science,
National Academy of Sciences of Ukraine, Vilde 5, 58001 Chernovtsy, Ukraine}
\affiliation{$^d$Max-Planck Institut f\"ur Mikrostrukturphysik,
Weinberg 2, 06120 Halle, Germany}
\affiliation{$^e$Department of Physics, Adam Mickiewicz University,
Umultowska 85, 61-614 Pozna\'n, Poland}
\affiliation{$^f$Institute of Molecular Physics, Polish Academy of Sciences,
Smoluchowskiego 17, 60-179 Pozna\'n, Poland}

\date{\today }

\begin{abstract}
The transmission  of correlated electrons through  a domain wall  in
ferromagnetic quasi-one-dimensional systems is studied
theoretically in the case when the domain wall width is comparable
with the Fermi wavelength of the charge carriers. The wall gives
rise to both potential and spin dependent scattering. Using a poor
man's renormalization group approach, we obtain scaling equations
for the scattering amplitudes. For repulsive interactions, the
wall is shown to reflect all incident electrons at the zero
temperature fixed points. In one of the fixed points the wall
additionally flips the spin of all incident electrons, generating
a finite spin current without associated charge current.
\end{abstract}

\maketitle
%%%%%%%%%%%%%%%%%%%%

{\it Introduction} Electronic properties of magnetic wires with
domain walls (DWs) attract much interest because of possible
applications of the associated magnetoresistance effect
\cite{wolf01}. So far, the puzzlingly huge magnetoresistance up to
thousands of percents in Ni wires and structured magnetic
semiconductors is not well understood
\cite{garcia99,chopra02,ruster03}. Most of the existing theories
of the DW resistance do not take into account the electron
correlations
\cite{tatara97,gorkom99,tagirov02,zhuravlev03,dugaev03}, except
for the case of a wide  DW (adiabatic regime) \cite{pereira04}.
However, in effectively one-dimensional (1D) structures one should
take into account the tendency to the non-Fermi-liquid behavior
\cite{solyom79}, possible enhancement of the
electron-impurity interaction \cite{kane92} and of the 
localization effects \cite{altshuler85}.
Moreover, non-adiabatic DWs (smaller or comparable to the Fermi wavelength)
% (even atomically sharp)
can be  achieved in ferromagnetic semiconductor wires
with constrictions.

In this Letter we study the effect of electron correlations on the
resistance of a ferromagnetic wire with a non-adiabatic DW. Since
the DW in a 1D wire acts as a localized spin-dependent scattering
center, a strong influence of electron correlations  is expected
\cite{kane92}. We apply the renormalization method
\cite{matveev93,devillard05} for the logarithmically diverging
terms in the perturbation theory of the electron interactions. We
find the zero-temperature fixed points for repulsive interactions,
in which the wall reflects all incident electrons. This might
explain the giant magnetoresistences observed in magnetic
nanowires. Moreover, we show that for one fixed point the electron
spin is reversed at the reflection. This leads to a nonzero spin
current with no charge current, which is of high interest for
applications in spintronics devices.

{\it Model and method} We consider a magnetic 1D system with a
local exchange coupling between conduction electrons and a spatially
varying magnetization ${\bf M(r)}$. The wire itself defines the
easy axis ($z$-axis), and a DW centered at $z=0$ separates two
regions with opposite magnetizations:  $M_z(z\rightarrow \pm
\infty) = \pm M_0$. Assuming  ${\bf M(r)}$ to lie in the $xz$
plane and the DW width to be smaller than the Fermi wavelength,
one can write the single-particle Hamiltonian as ($J>0$):
\begin{equation}
\hat H_0 = -\frac{\hbar^2}{2m} \frac{d^2}{dz^2} + \hbar V \delta(z)
+J M_z(z)\hat \sigma_z + \hbar \lambda   \delta(z)\hat \sigma_x\,,
\label{model}
\end{equation}
where the term $\hbar \lambda \delta(z)\hat \sigma_x$ describes
spin dependent scattering due to the $M_x(z)$ component
\cite{dugaev03}, $\hbar\lambda = J\int_{-\infty}^{\infty}
M_x(z)dz$, and $V$ is a potential (spin independent) scattering
term. We assume that spin-$\up$ electrons are spin majority ones
for $z<0$ and spin-minority for $z>0$.

An incident electron from the left with  momentum $k$ and spin
$\up$ (or $\dn$) can be transmitted to the $z>0$ region preserving
its spin, but changing momentum from $k$ to $k^-$ (or $k^+$),
given by $k^\pm=\sqrt{k^2 \pm 4 JM_0m/\hbar^2}$. If the
transmission occurs with spin reversal, the momentum $k$ remains
unchanged. The reflection amplitudes for spin-$\sigma$ electrons
with or without spin reversal are denoted by $r'_\sigma$ and
$r_\sigma$, respectively. The  same convention applies to the
transmission amplitudes $t'_\sigma$ and $t_\sigma$. The scattering
amplitudes are given by:
\begin{eqnarray}
t_{\up(\dn)}(k) &=& \frac{2\left( v+v^\mp + 2iV\right) v }
{ (v+v^\mp + 2iV)^2
+ 4\lambda^2} = r_{\up(\dn)}(k) + 1\,,\label{tup}\\
t_{\up(\dn)}'(k) &=& \frac{4i\lambda v}{(v+v^\mp + 2iV)^2 + 4\lambda^2}
= r_{\up(\dn)}'(k)\,,\label{tupl}
\end{eqnarray}
with $v=\hbar k/m$, $v^{\pm} = \hbar k^{\pm}/m $, where the upper (lower)
sign refers to $\up$ ($\dn$). We shall henceforth denote by
$\epsilon(\pm p, \sigma)$ the energy of a scattering state with
momentum $+p$ (or $-p$) and spin $\sigma$, incident from the left
(or right). The scattering amplitudes satisfy general relations
that can be found from a generalization of the Wronskian theorem
\cite{messiah, eu}  to spinor wave-functions: the Wronskian of any
two states having the same energy is a constant,
\begin{equation}
W(\psi_1, \psi_2)\equiv \psi_1^t(z) \; \frac{d\psi_2}{dz}
-\frac{d\psi_1^t}{dz}\; \psi_2(z)=const,
\label{Wronsk}
\end{equation}
where $\psi^t$ denotes the transpose of the spinor $\psi$.

In order to deal with the electron interactions, it is convenient
to rewrite the scattering states in second quantization form,
making use of right ($\hat a_{q\sigma}$) and left ($\hat
b_{q\sigma}$) moving plane wave states. We introduce operators
$\hat c_{k, \sigma}$ and $\hat d_{k, \sigma}$  for the eigenstates
corresponding to electrons incident from the left and right,
respectively. The plane wave operators are linear combinations of
the operators of scattering states. Electron interactions are then
introduced through the Hamiltonian
\begin{eqnarray}
\hat H_{int} &=& g_{1,\alpha,\beta}\int \frac{dk_1dq}{(2\pi)^2}
\hat a^\dagger_{k_1,\alpha} \hat b^\dagger_{k_2,\beta}
\hat a_{k_2+q,\beta} \hat b_{k_1-q,\alpha}\nonumber\\
&+&  g_{2,\alpha,\beta}\int \frac{dk_1dq}{(2\pi)^2} \hat
a^\dagger_{k_1,\alpha} \hat b^\dagger_{k_2,\beta} \hat
b_{k_2+q,\beta} \hat a_{k_1-q,\alpha} \,, \label{hint}
\end{eqnarray}
where the Greek letters denote spin indices, and the summation
convention over repeated indices is used. The coupling constants
$g_1$ and $g_2$ describe back and forward scattering processes
between electrons moving in opposite directions, respectively.
Because the Fermi momentum is spin dependent, we distinguish
between $g_{1(2)\up}$, which describe interaction between
spin-majority particles (that is spin-$\up$ on the left and
spin-$\dn$ on the right of the barrier) and $g_{1(2)\dn}$, which
describe interaction between spin-minority particles.
% (that is spin-$\dn$ on the left and spin-$\up$
%  on the right of the barrier).
We use $g_{1(2)\perp}$ to denote interaction between
particles with opposite spin. The forward scattering process
between particles which move in the same direction will not affect
the transmission amplitudes, although it will renormalize the
Fermi velocity \cite{solyom79,matveev93}. This effect is equivalent to an
effective mass renormalization and the electrons with different
spin orientations may turn out to have different effective masses,
in which case our  calculations remain valid \cite{eu}.

Following Ref.~\cite{matveev93}, the corrections to the
transmission amplitudes are calculated to first order in the
perturbation $\hat H_{int}$. Such corrections  diverge
logarithmically  near the Fermi level and  will  be dealt with in
a poor man's renormalization approach. The perturbative correction
to $t_\sigma(p')$ can be obtained from the perturbation expansion
for the Matsubara propagator
\begin{equation}
{\cal G}_\sigma(\tau) =
 -\la T_\tau e^{-\int \hat H_{int}(\tau')d\tau'}
 \hat a_{p, \sigma} (\tau) \hat c_{p',\sigma}^\dagger\ra _0\,,
\label{G}
\end{equation}
where $\la ... \ra_0$ denotes the average in the Fermi sea {\it of
scattering states}.
 The zero-order propagator  for $\sigma=\up$  is:
\begin{equation}
{\cal G}_\up^{(0)}(i\omega)  = \frac{1}{i\omega - \epsilon(p',
\up)}\left[ \frac{i}{p-p'+i0} - \frac{i t_\up(p')}{p-p'^--i0}
\right]\,, \label{zero}
\end{equation}
where  $0$ denotes a positive infinitesimal. The poles in the
denominators identify the semi-axis on which the electron behaves
as a right moving plane wave. The transmission amplitude appears
associated with the denominator $p-p'^--i0$ which, for the
variable $p$, gives a pole in the upper half plane. The  meaning
of this pole is that the transmitted particle is right-moving in
the $z>0$ half-axis. Our strategy is to calculate the first order
correction term  to ${\cal G}$, in which
 a pole in $p-p'^--i0$ will appear with the residue
 $-i\delta t_\up(p')$, which is the transmission amplitude correction.

\begin{figure}[ht]
\begin{center}
\epsfxsize=8cm
\epsfbox{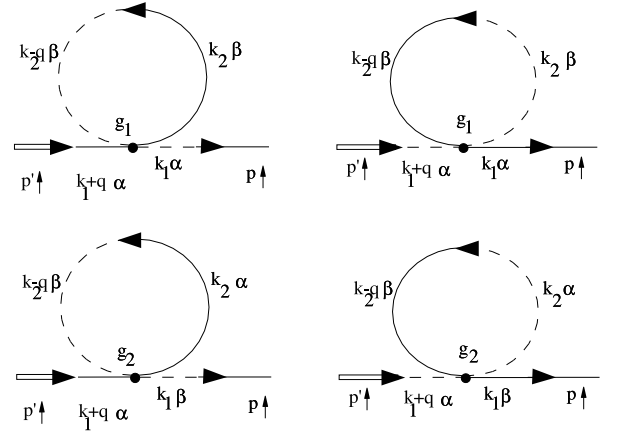}
\end{center}
\vspace*{-0.5cm}
\caption{Feynman diagrams for the first order contribution ${\cal
G}^{(1)}$ to the propagator (\ref{G}). The scattering state is
represented by a double line, the $\hat a$ ($\hat b$) particle is
represented by a continuous (dashed) line. The loop represents the
 Hartree-Fock potential of the Fermi
sea. The scattered electron exchanges momentum $q$ with the Fermi
sea.} \label{fig2}
\end{figure}

The diagrammatic  representation of ${\cal G}_\up^{(1)}$ is shown
in Fig.~\ref{fig2}. The horizontal lines represent the electron
scattered  by the Hartree-Fock potential of the Fermi sea. Only
processes where the electron is back-scattered by the Fermi sea
are logarithmically large \cite{matveev93}. Consider, for
instance, the upper left diagram --  an electron, initially in
state $ c_{p', \up}$ close to the Fermi level passes through the
DW as a right-moving ($\hat a$) particle. Then, it is reflected
(from $\hat a$ to $\hat b$ particle) while exchanging momentum $q$
with the Fermi sea on the $z>0$ semi-axis. Finally, it is
reflected by the DW again, becoming a spin-up right moving
particle of momentum $p$. A logarithmic divergence occurs if the
polarized Fermi sea can provide exactly the momentum that is
required to keep the electron always near the Fermi level in
the intermediate virtual steps. According to the physical
interpretation of the diagrams, we always know on which side of
the DW the interaction with the Fermi sea (closed loop in the
diagram) is taking place.
%%%%%%%%%%%%%%%%%%%%%%%%% FIGURE %%%%%%%%%%%%%%%%%%%%%%%%%%%%%%%%%%
%%%%%%%%%%%%%%%%%%%%%%%%% END FIGURE %%%%%%%%%%%%%%%%%%%%%%%%%%%%%%%

We use Fermi level velocities $v_\pm$ and wavevectors $k_{F\pm}$ for majority or
minority spin particles, henceforth.
It can be seen that $g_{1\perp}$ terms are proportional to
$\log|k_{F+}-k_{F-}|$, so they do not diverge. Logarithmic
divergence would be restored in a spin degenerate system
($k_{F+}=k_{F-}$). This can be understood from the diagrams in
Fig.~\ref{fig2} as follows: the electron with spin $\alpha$ is
reflected by a polarized Fermi sea with spin $-\alpha$. The
momentum provided by the Fermi sea is $2k_{F-\alpha}$, while the
momentum required by the electron is  $2k_{F\alpha}$. The
$g_{2\perp}$ terms  produce logarithmic divergences that would not
exist in the absence of spin-flip scattering ($t'=r'=0$). For the
calculation of $\delta t_\sigma'(p')$  the propagators we need to
consider are $ -\la T_\tau \hat a_{p, -\sigma}(\tau)  \hat c_{p',
\sigma}^\dagger\ra$.

The expression for  $\delta t_\up(p')$ is directly proportional to
$\log (|\epsilon'|/D)$, where  $\epsilon'$ denotes the energy of
the scattered electron and $D$ is an energy scale near the Fermi
level where the electron dispersion can be linearized [$\delta
t_\up(p')/\log (|\epsilon'|/D)$ is given by the right-hand side of
equation (\ref{tupdif}) below]. The logarithmically divergent
perturbation can be dealt with using a renormalization procedure
\cite{matveev93}; reducing  step by step the bandwidth $D$ and
removing states near the band edge is compensated by
renormalization of $t_\up$. Applying this procedure and noting
that $t_\up+ \delta t_\up$ remains invariant as $D$ is reduced,
one finds the following differential equation:
$$
dt_\up \ + \ \frac{\partial \, \delta t_\up}{\partial D}\ dD = 0\,.
$$
Now, we introduce the variable $\xi = \log (D/D_0)$, which is
integrated from $0$ to $\log (|\epsilon'|/D_0)$, corresponding to
the fact that the bandwidth is progressively reduced from $D_0$ to
$|\epsilon'|$ (which will be taken as temperature:
$|\epsilon'|=T$). It is convenient to rewrite the interaction
parameters as
$g_{\up(\dn)} =(g_{2\up(\dn)}-g_{1\up(\dn)})/4hv_{+(-)}$,
%$g_{\up} =(g_{2\up}-g_{1\up})/4hv_{+}$,
%$g_{\dn} =(g_{2\dn}-g_{1\dn)}/4hv_{-}$,
$g_\perp=g_{2\perp}/{2h(v_+ + v_-)}$.
The scaling differential equations for the transmission amplitudes
are
%%%%%%%%%%%%%%%%%%%%% t_\up :
\begin{eqnarray}
\frac{d t_\up}{d\xi} &=&
 g_\dn \left[ \   r_\dn^* r_\dn t_\up + r_\dn^* r_\up' t_\dn'\ \right]
%  \nonumber\\ &+&
+g_\up \left[ \  r_\up^* r_\up  t_\up + r_\up^*  r_\up' t_\up'\  \right] \nonumber\\
&+&
g_\perp\left[
r_\dn'^*  r_\up' t_\up\  +\
r_\up'^*  r_\dn t_\up'\
%\right. \nonumber\\
% &+& \left.
+ r_\up'^* r_\up t_\dn' + r_\dn'^* r_\up' t_\up\  \right] ,
\hskip0.5cm
\label{tupdif}
\end{eqnarray}
%%%%%%%%%%%%%%%%%%%%%% t_\up' :
\begin{eqnarray}
\frac{d t_\up'}{d\xi}
=2g_\dn  r_\dn^*  r_\dn' t_\up + 2g_\up r_\up^* r_\up t_\up'
%\nonumber\\ &+&
+2g_\perp \left[
r_\dn'^* r_\up t_\up\ + \
r_\up'^* r_\dn' t_\up'\ \right]\,.
%\hskip0.2cm
\label{tpupdif}
\end{eqnarray}

Equations for the reflection amplitudes $r_\sigma (p')$ and $r_\sigma' (p')$ can
be obtained from  the propagators $ -\la T_\tau \hat b_{p,
\pm\sigma}(\tau)  \hat c_{p', \sigma}^\dagger\ra$.
The equation for  $r_\up (p')$ is
\begin{eqnarray}
\frac{ d r_\up}{d\xi}
=g_\up\left[  r_\up^*  r_\up  r_\up\  +\  r_\up^* t_\up't_\up' \right]
% \nonumber\\ &+&
+g_\dn\left[  r_\dn^*  t_\up  t_\dn\  +r_\dn^* r_\dn'  r_\up'\
\right]\nonumber\\
+g_\perp\left[ r_\up'^* r_\dn' r_\up\
+r_\dn'^* r_\up' r_\up\
%\right. \nonumber\\ &+& \left.
+r_\up'^* t_\dn t_\up'\
+r_\dn'^* t_\up t_\up' \right]
-g_\up r_\up \hskip0.3cm
\label{ruplogs}
\end{eqnarray}
and the equation for  $r_\up' (p')$ is
\begin{eqnarray}
\frac{ d r_\up'}{d\xi}
=g_\up \left[ r_\up^*  r_\up  r_\up'\  +\ r_\up^* t_\up't_\up\ \right]
%  \nonumber\\ &+&
+g_\dn \left[ r_\dn^*  t_\up  t_\dn'\  + r_\dn^* r_\dn  r_\up'\  \right]
\nonumber\\
+g_\perp \left[ r_\up'^* r_\dn r_\up
+r_\dn'^* r_\up' r_\up'
%\right. \nonumber\\  &+& \left.
+r_\up'^* t_\dn' t_\up'\
+r_\dn'^* t_\up t_\up \right] -g_\perp r_\up' \,.
\hskip0.3cm
\label{rupup}
\end{eqnarray}
The scaling equations for spin-$\dn$ amplitudes
follow from the above by simply inverting the spin and velocity indices. Equations
(\ref{tupdif})-(\ref{rupup}) are the central result of this paper.
Theorem (\ref{Wronsk}) gives $v_-/v_+=t_\dn/t_\up=r_\dn'/r_\up'$.
A standard second-order renormalization group treatment shows that
in a 1D magnetized system the coupling constants in equation
(\ref{hint}) are not renormalized because the logarithmically
divergent contributions cancel each other.

{\it Fixed points} We have performed a numerical analysis of the
scaling equations using the DW model (\ref{model}), with $V=0$, 
for the initial parameters. 
We now analyze the nature of the fixed points
predicted by the scaling equations. The parameters of the model
which enter the scaling equations are $g_\up$, $g_\dn$, $g_\perp$,
and the ratio $v_-/v_+$. For repulsive interactions ($g_\up,
g_\dn, g_\perp>0$) the system flows to insulator fixed points. For
$\lambda/v_+$ larger than about $0.1$, all transmission amplitudes
vanish faster than any reflection amplitude as $T\rightarrow 0$.
In this limit we may rewrite the scaling equations  for $r_\up$,
$r_\up'$ and the theorem (\ref{Wronsk}) neglecting the small
transmission amplitudes. Theorem (\ref{Wronsk}) for
$W(\psi_{k,\up}^*, \psi_{k^-,\dn})$ tells us that $r_\dn^* r_\dn'
+ r_\up r_\dn'^* = 0$. The charge conservation condition is
satisfied solely by the reflections,
\begin{equation}
1\ =\ |r_\up|^2 + \frac{v_-}{v_+} |r_\up'|^2\  =\   |r_\dn|^2 + \frac{v_-}{v_+} |r_\up'|^2\,,
\label{a12ins}
\end{equation}
from which we conclude that $ |r_\up| =  |r_\dn|$ at the fixed
point. Equations (\ref{ruplogs}) and  (\ref{rupup})  now take the
form
\begin{eqnarray}
\frac{ d r_\up}{d\xi}
=\frac{v_-}{v_+} \  \left(\ 2g_\perp
-g_\up-g_\dn  \right)  \left(\  1-|r_\up|^2 \right) \ r_\up\,,
\label{finalrup}
%\end{eqnarray}
\\
%\begin{equation}
\frac{ d r_\up'}{d\xi} = \left(\  g_\up + g_\dn - 2 g_\perp\right)
 \left( 1-\frac{v_-}{v_+} |r_\up'|^2 \right) r_\up'\,.
\label{ruppfinal}
\end{eqnarray}
In the derivation of (\ref{finalrup}) and (\ref{ruppfinal})
only the smallness of  the transmissions amplitudes was assumed.
 The phases of
the complex numbers $r_\up$,  $r_\up'$ are unchanged during
scaling. The two fixed points we may consider correspond to
$r_\up$ approaching 0, or $|r_\up|$ approaching 1.

The situation  $|r_\up|\rightarrow 0$ requires $2g_\perp -
g_\up-g_\dn >0$ and, by charge conservation we have
$|r_\up'|\rightarrow \sqrt{v_+/v_-}$. Upon integrating
(\ref{ruppfinal}) with $\xi$ going from $0$ to $\log(T/D_0)$, the
amplitude $r_\up'$ will vary from $r_{\up,0}'$ to $r_\up'(T)$.
Introducing the   reflection coefficient ${\cal R}_\up' =
(v_-/v_+) |r_\up'|^2  $, we obtain
\begin{equation}
{\cal R}_\up'(T)=\frac{\frac{{\cal R}_{\up,0}'}{1-{\cal R}_{\up,0}'}
\left(\frac{T}{D_0}\right)^{2\left(g_\up+g_\dn-2g_\perp\right)}}
{ 1+ \frac{{\cal R}_{\up,0}'}{1-{\cal R}_{\up,0}'}
\left(\frac{T}{D_0}\right)^{2\left(g_\up+g_\dn-2g_\perp\right)}}\,.
\label{sfins}
\end{equation}
If  $2g_\perp - g_\up-g_\dn >0$ then ${\cal R}_\up'(T)\rightarrow
1$ as $T\rightarrow 0$. The DW reflects all incident electrons
while additionally reversing their spin -- it is a $100\%$
``spin-flip reflector'' at zero temperature, generating a finite
net spin current but no charge current. In order to find the low
$T$ behavior of transmissions we put $r_\sigma=0$, $|r_\up'|=\sqrt{v_+/v_-}$
 in the scaling equations for the transmission amplitudes and obtain
$|t_\up| \sim |t_\up'|\sim |t_\dn'| \sim T^{2g_\perp}$.

In the regime where $g_\up+g_\dn-2g_\perp >0$ we have ${\cal
R}_\up'(T)\rightarrow 0$,  ${\cal R}_\up(T)\rightarrow 1$. The DW
reflects then all incident electrons while  preserving their spin.
The scaling  equations for the transmission amplitudes yield
$|t_\up| \sim T^{g_\up + g_\dn}$,  $|t_\sigma'| \sim
T^{2g_\sigma}$.

If $g_\up+g_\dn-2g_\perp =0$ then both  $r_\sigma'(T)$ and
$r_\sigma(T)$ tend to finite values. The scaling equations for
$t_\up$, $t_\sigma'$,  with constant reflection amplitudes,
become a linear algebraic 3 by 3 system. The eigenvalues of the
matrix give the temperature exponents and each transmission
amplitude will be a linear combination of the three powers of $T$.
For decreasing temperature there may be crossovers from one
exponent to the other and the lowest one dominates as
$T\rightarrow 0$.

For smaller values of $\lambda/v_+$ ({\it i.e.,} smaller than
about $0.1$) in the Hamiltonian (\ref{model}), the system flows to
a fixed point  where $r_\up'$ vanishes about as fast as the
transmissions and $|r_\sigma|\rightarrow 1$.  The transmission
amplitudes still scale to zero as $|t_\up| \sim T^{g_\up +
g_\dn}$,  $|t_\sigma'| \sim  T^{2g_\sigma}$. The  scaling equation
for $r_\up'$ can be linearized in $r_\up'$. Neglecting the second
order terms in  $t$, $t'$, and  considering that $|r_\sigma
|\rightarrow 1$, we obtain $|{\cal R}_\up'| \sim T^{2\left(g_\up +
g_\dn\right)}$. In the scaling equation (\ref{ruplogs}) for
$r_\up$ we cannot neglect the terms containing  transmission
amplitudes on the right hand side. The behavior of $r_\sigma$ as
$T\rightarrow 0$ can be found from the charge conservation
condition:
 $1- |r_\sigma|^2 \sim T^{{\rm min}
 \left\{2(g_\up + g_\dn), 4g_\sigma \right\}}$.

We may estimate the $\lambda$ parameter in Eq.(\ref{model}) by
assuming that $\vec{M}(z)= M_0 \cos\theta(z)\hat{\vec{ z}} + M_0
\sin\theta(z){\hat{\vec{ x}}}$ with $\cos\theta(z) = \tanh(z/L)$,
 where $L$ is the length of the DW \cite{pereira04}.   We then
have $ \lambda   = \pi M_0 L/\hbar$, implying that
\begin{eqnarray}
\frac{\lambda}{v_+} = \pi m\frac{JM_0} {\hbar^2 k_{F+}^2} (L
k_{F+}) . \label{lam}
\end{eqnarray}
The condition for  the DW to be  smaller than the Fermi wavelength
is $L k_{F+} <2\pi$. The smaller Fermi wavelength is that of
spin-majority electrons, $2\pi/k_{F+}$. The ratio  $v_-/v_+$
depends on the degree of polarization of the electron system. In a
1D nonmagnetic system there is a single Fermi momentum, $k_F$, for
up and down electrons and a Fermi energy $E_F=\hbar^2k_F^2/(2m)$.
Once the system becomes magnetized, there is a  Zeeman energy
shift of the bands, $\Delta E/2 = JM_0$, and the two new Fermi
momenta, $k_{F\pm}$, satisfy
\begin{eqnarray}
\frac{k_{F\pm}}{k_F} = 1\pm \frac{\Delta E}{4E_F}\, .
\end{eqnarray}
Inserting this result in Eq.(\ref{lam}) above, we obtain
\begin{eqnarray}
\frac{\lambda}{v_+} =\pi \frac{(\Delta E/4E_F)}{\left[ 1 + (\Delta
E/4E_F)  \right]^2}  \ (L k_+)\ . \label{estima}
\end{eqnarray}
In the full polarization limit  $k_{F-}=0$, $k_{F+}=2k_F$, and
Eq.~(\ref{estima}) gives $\lambda /v_+ \approx 0.79\, L k_{F+}$.
Typical values for a non-fully polarized system are  $ E_F =
90\,$meV and  $\Delta E=30\, $meV  \cite{ruster03}. In this case
we have  $v_-/v_+=0.84$ and  Eq.~(\ref{estima}) gives $\lambda
/v_+ \approx 0.22  L k_{F+}$. Therefore, if $L k_{F+}$ is smaller
than about $2\pi$, the system can flow to any of the fixed points
described above.

Lateral quantization may produce several channels. The possibility
of  inter-channel scattering then arises due to two
causes: {\it (i)} electron  interactions (which would
require a modification of our theory to allow for inter-channel
scattering); {\it (ii)} impurity scattering. For the latter
to be negligible  the electron mean free path  must be larger
than the size of the constriction pinning the DW.

The above spin-flip reflector DW was not found in Ref
\cite{pereira04}. This is because the adiabatic DW considered in
Ref \cite{pereira04} is a poor spin-flip reflector at the
noninteracting level already -- as in the regime of small
$\lambda/v_+$ above.

%%%%%%%%%%%%%%%%%%%%%%%%%%%%% Acknowledgements %%%%%%%%%%%%%%%%%%%%%
%\section*{Acknowledgments}
Discussions with P. A. Lee, A. H. Castro Neto and P. Sacramento
are gratefully acknowledged. M.A.N.A. is grateful to
Funda\c{c}\~ao para a Ci\^encia e Tecnologia for a sabbatical
grant. This research was supported by Portuguese program POCI
under Grant No. POCI/FIS/58746/2004, EU through RTN Spintronics
(contract HPRN-CT-2000-000302), Polish State Committee for
Scientific Research under Grant No. 2~P03B~053~25, and by STCU
Grant No.~3098 in Ukraine.
%%%%%%%%%%%%%%%%%%%%%%%%%%%%%%%%%%%%%%%%%%%%%%%%%%%%%%%%%%%%%%%%%%%%

\end{document}